\documentclass[journal]{IEEEtran}

\usepackage[english]{babel} 
\usepackage{amsmath,amsfonts,amsthm} 
\usepackage[utf8]{inputenc}
\usepackage{array}
\usepackage{float}
\usepackage{graphicx}
\usepackage[justification=centering]{caption}
\usepackage{array,booktabs}
\usepackage{csquotes}
\usepackage{refcount}
\usepackage{url}
\usepackage{listings}
\usepackage[table,xcdraw]{xcolor}
\usepackage{tabularx,ragged2e}
\usepackage{graphics}
\usepackage{pbox}
\usepackage[section]{placeins}
\lstdefinestyle{xmllistingstyle}{
    language=xml,
    frame=single,
    rulesepcolor=\color{black},
    xleftmargin=0pt,
    framexleftmargin=2pt,
    keywordstyle=\color{blue}\bf,
    commentstyle=\color{teal},
    stringstyle=\color{black}\ttfamily\bfseries,
    captionpos=b,
    numbers=none,
    numbersep=2pt,
    breaklines=true,
    showstringspaces=false,
    basicstyle=\scriptsize,
    escapeinside={(*@}{@*)}
    }

\ifCLASSOPTIONcompsoc
  \usepackage{array,booktabs}
  \usepackage{url}
\else
  \usepackage{array,booktabs}
  \usepackage[table,xcdraw]{xcolor}
  \usepackage{cite}
\fi

\begin{document}
\title{I Probe, Therefore I Am: Designing a Virtual Journalist with Human Emotions}

\author{
		Kevin K. Bowden,~\IEEEmembership{University of California, Santa Cruz,~}
       Tommy Nilsson,~\IEEEmembership{~University of Nottingham,}
        Christine P. Spencer,~\IEEEmembership{Queen's University Belfast,}
        K\"{u}bra Cengiz,~\IEEEmembership{~Istanbul Technical University,}
        Alexandru Ghitulescu,~\IEEEmembership{~University of Nottingham}
        and~Jelte B. van Waterschoot,~\IEEEmembership{~University of Twente~}

\IEEEcompsocitemizethanks{\IEEEcompsocthanksitem K. K. Bowden is with the Natural Language and Dialog Systems Lab, University of California, Santa Cruz, USA. Email: kkbowden@ucsc.edu.
\IEEEcompsocthanksitem T. Nilsson is with The Mixed Reality Laboratory, University of Nottingham, UK, and The National Institute of Mental Health, Klecany, Czech Republic. Email: psxtn2@nottingham.ac.uk.\IEEEcompsocthanksitem C. Spencer is with the Social Interactions Lab, Queen's University Belfast,
\protect\\
E-mail: cspencer03@qub.ac.uk
\IEEEcompsocthanksitem K. Cengiz is with the Istanbul Technical University, Turkey. Email: ce.kubra@gmail.com.
\IEEEcompsocthanksitem A. Ghitulescu is with the Computer Vision Laboratory, University of Nottingham, UK. Email: alex.ghitulescu@gmail.com.
\IEEEcompsocthanksitem J. B. van Waterschoot is with the Human-Media Interaction Lab, University of Twente, Netherlands. Email: jbvanwaterschoot@utwente.nl.

}

\thanks{Manuscript received Month day, year; revised Month day, year.}}

\markboth{Journal of \LaTeX\ Class Files,~Vol.~14, No.~8, August~2015}%
{Shell \MakeLowercase{\textit{et al.}}: The Virtual Human Journalist}

\maketitle

\pagestyle{empty}
\thispagestyle{empty}

\begin{abstract}
By utilizing different communication channels, such as verbal language, gestures or facial expressions, virtually embodied interactive humans hold a unique potential to bridge the gap between human-computer interaction and actual interhuman communication. The use of virtual humans is consequently becoming increasingly popular in a wide range of areas where such a natural communication might be beneficial, including entertainment, education, mental health research and beyond. 
Behind this development lies a series of technological advances in a multitude of disciplines, most notably natural language processing, computer vision, and speech synthesis. In this paper we discuss a Virtual Human Journalist, a project employing a number of novel solutions from these disciplines with the goal to demonstrate their viability by producing a humanoid conversational agent capable of naturally eliciting and reacting to information from a human user. 
A set of qualitative and quantitative evaluation sessions demonstrated the technical feasibility of the system whilst uncovering a number of deficits in its capacity to engage users in a way that would be perceived as natural and emotionally engaging. We argue that naturalness should not always be seen as a desirable goal and suggest that deliberately suppressing the naturalness of virtual human interactions, such as by altering its personality cues, might in some cases yield more desirable results. 
\end{abstract}


\IEEEpeerreviewmaketitle

\section{Introduction}
\IEEEPARstart{T}{he} principal goal of the EU’s ARIA-VALUSPA\footnote{ARIA-VALUSPA is short for Artificial Retrieval of Information Assistants – Virtual Agents with Linguistic Understanding, Social skills, and Personalised Aspects. Full detail of the project can be found at http://aria-agent.eu/ \label{MyFootNoteLabel}} project is to develop a framework for the production of virtual humans capable of engaging in a multi-modal speech-based social interaction with a user in a variety of situations. 

In comparison with text-based interactions, speech-based interactions are considered to have a number of well-documented advantages. It has, for instance, been established that direct, first-person speech increases the memorability of a message \cite{tannen1989, schiffrin1981}. Moreover, research also shows that dialogue is more persuasive than monolog \cite{piwek2007}. To achieve a natural conversational flow, it is necessary for virtual humans to be able to sustain an interaction whilst reacting appropriately to both the verbal and non-verbal behavioral cues of the user. This demands a range of diverse technological solutions to be in place and operate in unison. 

For this purpose, the ARIA-VALUSPA project brings together an international consortium of academic and industrial institutions, including CereProc\footnote{CereProc, headquartered in Edinburgh, UK, is a leading developer of text-to-speech solutions. Full company overview can be found at https://www.cereproc.com/en/ \label{MyFootNoteLabel}} specialized in the development of  text-to-speech (TTS) solutions, a research team from Paris Telecom focusing on the development of virtually embodied human characters and a number of other academic research laboratories across the UK, France, Germany and the Netherlands. In other words, the project builds on a broad and interdisciplinary research area, spanning a multitude of fields, including psychology, linguistics, artificial intelligence and computer graphics. 

By summer 2016 the project had progressed to a stage where a prototype facilitating a relatively successful interaction between the user and a virtual human could be produced. To demonstrate the viability of the technology, a spin-off project called Virtual Human Journalist was carried out during the 2016 eNTERFACE workshop. A small team of international students were given access to the technical solutions developed by the ARIA-VALUSPA consortium and tasked to produce a virtual human journalist capable of interviewing and eliciting information from a user. This information was to be stored, processed, and eventually relayed back to the user in a coherent, ostensibly authentic and contextually-appropriate way. In other words, the goal was not simply to develop a working technical solution, but also to produce an experience that would be perceived by users as natural and believable through the demonstration of responsiveness. 

Over the course of four weeks, the team developed a dialogue management system centered around a set of probing questions designed to elicit a maximum amount of relevant information from the user. Informal qualitative evaluations of the agent's conversational performance were continuously carried out in parallel to the development of the dialogue system in order to assess its performance in a real world setting. Besides assessing the perceived quality of the virtual journalist's verbal and non-verbal behavior, a set of quantitative evaluations were likewise aimed at assessing the degree of user emotional engagement and the likability of the agent induced by the system.  The overall aim of the workshop was therefore to design and test the efficacy of the system to elicit information from the user whilst maintaining user interest and emotional engagement. This was evaluated by the agent's ability to sustain at least a five minute interaction with the user by using a number of information-eliciting strategies. User interest and emotional engagement were assessed via a post-interaction questionnaire.

A virtual human journalist possessing the ability to successfully elicit information in a sustained and naturally flowing interaction whilst appropriately addressing the user's emotional state may offer multiple valuable future applications. For example, the development of an empathic embodied agent which demonstrates an interest in learning more about the user could offer social companionship to elderly users, or members of vulnerable social groups. In terms of other practical applications, for example, such an agent could reduce labor-intensive human activities by successfully eliciting and storing user's information into a database via a social interaction conducted by telephone. 

The following section of this paper provides a brief overview of key theoretical concepts relevant to our work. The third section describes the system design of the virtual human journalist. The fourth section describes the evaluation of the agent, containing two small pilot studies for data collection and testing. In the fifth section we describe some of the limitations of our project. In the sixth section we describe alternative means for improving our system design. Finally, in the seventh section we make concluding remarks about developing an engaging embodied virtual agent for information extraction.

\section{Background}

A critical factor in interpersonal interactions and relationships is empathy and emotional responsiveness, \cite{decety2006} due to its ability to help individuals to recognize, understand, predict and respond appropriately to the behavior of others \cite{vignemont2006}. If virtual humans are to achieve interaction quality which is similar, if not equal to, human-human communication, developing convincing simulation of empathic abilities will be crucial. Haase and Tepper \cite{haase1972} suggest that both verbal and non-verbal cues play an important role in the complex multi-channel process of establishing a sense of emotional responsiveness. 

Believable empathic cues in virtual humans can consequently come in various forms. In a multi-modal interface, a user might for example perform gestures, such as pointing at literal or figurative objects whilst speaking. It is critical for virtual humans to be able to accurately recognize and "understand" the emotional relevance of such gestures in a particular social context, as well as possessing its own contextually-appropriate gesturing abilities. 

Other crucial non-verbal cues include facial expressions. Of particular relevance are the attempts of psychologists to develop facial expression classification systems \cite{fast2016, shaver1987}. For instance, Phillip et al. identified 135 emotion names, which the authors clustered hierarchically \cite{shaver1987}. These taxonomies provide us with a broad basis for the mapping of emotion to verbal language, and in turn the opportunity to accompany it with convincing non-verbal signals. The viability of such emotional models has been demonstrated by their successful implementation in video games. \cite{andry2016}

Similarly, various back-channeling signals can indicate to an interaction partner that one is emotionally engaged in an interaction. For example, back-channeling behaviors are thought to demonstrate engaged listenership \cite{lambertz2011}, whilst increased behavioral mimicry can signal greater rapport and an increased desire for affiliation \cite{lakin2003}. It therefore follows that, if human-computer interactions are to effectively simulate human-human communication, the agent must display a level of emotional responsiveness and full repertoire of verbal and non-verbal behaviors to be an engaging interaction partner \cite{mckeown2015}.

We centered the virtual human journalist system around a specific knowledge domain from which it can retrieve information pertaining to a particular subject area. The virtual human is also capable of prompting the user to request more information about a pre-defined subject, by asking questions such as "Would you like me to tell you about \textit{X}?", where \textit{X} is an example taken from a list of pre-defined subjects. However, the agent's conversational abilities are currently limited by the constraints of the domain. Given that the goal is to engage the user in a sustained and smooth-flowing interaction, this limitation constitutes a problem. The current conversational dynamic is rather one-sided, and subsequently, less naturalistic and representative of human-human communication. It thus appeared that in order to progress from simply an information-retrieval assistant to a plausible and engaging conversational partner, it would be necessary for the virtual human to demonstrate a more well-defined knowledge base. Moreover, this knowledge base ought also be dynamically expandable with additional information retrieved about the user through a series of probing questions. It was therefore seen as desirable to enhance the existing capabilities of the virtual human by integrating various information-eliciting strategies into its design. Such information-eliciting strategies could involve a range of elements making interaction with the virtual human feel more natural and engaging, including personality cues and simulation of empathy. In other words, it was seen as imperative to design a virtual human in a way that would allow it to engage the user and maintain user interest by demonstrating responsiveness to the user's emotional state. We hypothesize that improving the information gathering in this manner should in turn help expand the knowledge base so that more information could be processed and ultimately relayed back to the user. We predict these abilities will demonstrate a sense of higher cognition and a more sophisticated level of understanding. 

\section{System Design}

\begin{figure}[h!]
\begin{center}

\includegraphics[scale=0.4]{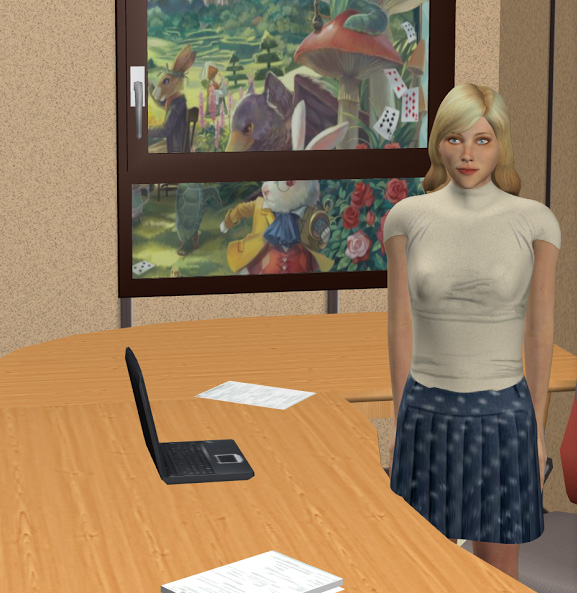}
\caption{Alice - the virtual human journalist}
\label{greta}
\end{center}
\vspace{-10pt}
\end{figure}

Here we describe the implementation of the virtual journalist. We start by elaborating on existing components and how we extended the virtual agent from the ARIA-VALUSPA project to become a virtual journalist. We will close this section with an example of a dialogue and how the pipeline processes the user input in this example. ``Alice'' was the agent selected for use as the interviewer, adopted from the ARIA-VALUSPA project\footnote{\url{https://github.com/ARIA-VALUSPA/ARIA-System}}. The agent understands spoken natural language and non-verbal behavior and uses both verbal and non-verbal behavior to converse with users. Alice's visual appearance is displayed in Figure \ref{greta}. 

\begin{figure*}
\begin{center}
\includegraphics[scale=.7]{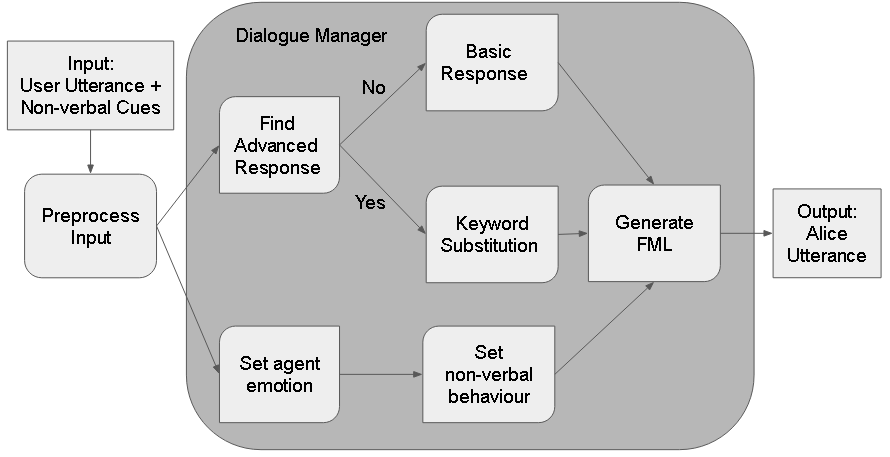}
\caption{Our general system pipeline}
\label{pipeline}
\end{center}
\vspace{-10pt}
\end{figure*}

\subsection{Preprocessing}

The input is first preprocessed by implementing components from the Stanford Natural Language Parser (NLP) toolkit \cite{stanford}. Specifically we used the dependency parser and subsequently the part of speech tagger to determine the structure of the sentence. We identify terms as important based on their part of speech and syntactic role within a sentence. We create nodes in a hand crafted knowledge base for important nodes within the sentence such as the named entities, proper nouns, and the subject. Additionally, we track the attributes which modify a node, possessions a node claims ownership over, and the other names a node can be referred to as by using light weight handcrafted anaphoric resolution. We also attempt to identify nodes which appear to be related to each other based on frequent co-occurrence in the same context.

\subsection{Dialogue Manager}
Flipper served as the dialogue manager (DM), structuring the dialogue dynamically between the agent and the user, based on an information-state approach to dialogues \cite{ter2011flipper}\cite{traum2003information}. Flipper does this by keeping track of the verbal (e.g. subject, proper nouns) and non-verbal (e.g. emotion) input from the user in the IS \cite{ter2011flipper}. Based on certain keywords and non-verbal user behavior defined in Flipper's templates, the agent would formulate its intent, i.e. how it wants to respond to the user. Every such possible response was written manually using the Functional Markup Language (FML), each defining the appropriate semantic units for the specific communicative intent \cite{heylen2008next}\cite{bunt2010towards}. The advantage of using FML is that the dialogue manager only has to care about selecting an agent intent, while the behavior planner fills in the blanks for how exactly to do perform the intent. For example in the DM you can define to say something angrily, and the behavior planner will decide if this is expressed via facial expressions or changing the pitch. The workings of the behavior planner are worked out in the next section. 

\begin {table}[H]
\begin{center}
\begin{tabular}{|p{8cm}|}
\hline
\smallskip
Hey, my name is Alice, what's your name?  \\
\smallskip
Nice to meet you \{name\}. So what do you do for fun in your spare time?  \\
\smallskip
So what do you specifically like about \{X\}? \\
\smallskip
So, you spoke a bit about \{Y\}. Why don't you tell me more about \{Y\}?\\
\smallskip
Nice, OK. Maybe we can chat about something else now. Do you have any pet?\\
\hline
\end{tabular}
\caption {Example of conversational questions used by Alice}
\end{center}
\vspace{-5mm}
\end {table}

We formulated questions of which some were emphatically phrased, based on findings of our first pilot study. Questions were designed to elicit usable nouns of the user, whilst being general enough to be applicable to most contexts, such as ``\textit{What do you do for a living?}".  Each subsequent response would then be dependent on the user's input. For example, if the user provided a long, informative answer, the agent could ask a few follow-up questions until the topic was exhausted. Conversely, if the user did not provide an adequate amount of information about a topic (\textit{X}), the agent could ask a question prompting more elaboration, such as ``\textit{I see. I'd be very interested in hearing more about X. Could you tell me a little bit more about that?"}. If the user's response was too complex, the agent then asked the user to rephrase the information. 

Finally, if the conversation began to run dry, or the topic became exhausted, the agent asked a question prompting a new topic. The speech content was additionally categorized by emotional valence, in order for the agent to adequately address the emotional content of the user's response with appropriately timed facial expressions and gestures for displaying empathy. Each state was also linked to a pool of responses such that the system can trigger the correct state while also avoiding exact repetitions. Examples of probing questions are included in Table I.


In this stage we augment Flipper's ability to pick an appropriate response using the knowledge base we built in the preprocessing phase. We use Equation \ref{eq:context}, to determine how relevant a node is to the current context. Specifically, we account for three node facets. We are interested in the frequency of a node - the number of mentions within the conversation, the time since it's been last mentioned, and the preference of a node. Here we establish the preference of a node by examining the sentiment of the content for which it's been applied. Preference is represented on a scale of 0-1 where our system sets the default preference of a given node to be neutral with a score of .5. We leave the robustness of this feature for future work.   

\begin{equation} \label{eq:context}
	score(u) = (freq(u) - \frac{timeSinceLast(u)}{1000})pref(u)     
\end{equation}

Flipper determines if there is a state which we can transition to that leverages our knowledge with a high scoring utterance. If we find such a state, we perform the necessary substitutions to make a valid and relevant response. If we are unable to leverage any of our information, Flipper uses its default keyword matching to determine the next state or to switch to a new topic generically. This default keyword matching also takes into account synonyms from a handcrafted list via the original ARIA-VALUSPA project. The DM is also responsible for keeping the conversation moving, if the user hasn't responded in a set amount of time we'll prompt the user to continue the conversation. 

\label{sys:output}
\subsection{Output}

\begin{figure}[h!]
\begin{center}
\includegraphics[scale=0.50]{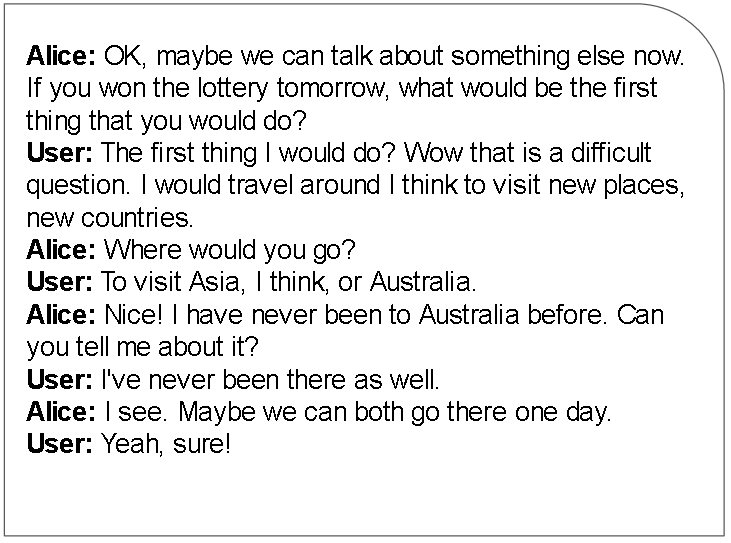}
\caption{An example conversation with Alice}
\label{auc1}
\end{center}

\vspace{-10pt}
\end{figure}

After the agent response was selected, Alice executed the behavior based on FML \cite{niewiadomski2009greta}. The FML could contain parameters such as where to apply a pitch accent or which emotion the agent has to show. These parameters can be loaded from the IS of the DM (e.g. deciding to mirror the emotion of the user) or be fixed in the FML-files (e.g. a question that always displays `surprise' behavior). An example conversation with Alice is included in Figure \ref{auc1}.

\section{Evaluation of the agent}
\begin{table}
 \begin{center}
  \begin{tabular}{ | c | c | c | }
    \hline
    Metric & $Mean_{1}$ & $Mean_{2}$  \\ \hline
    Liked Agent & 2.75 & 3   \\ \hline
    Likes Agents & 3.38 & 3.72   \\ \hline
    Engaging & 2.75 & 3.14   \\ \hline
    Enjoyment & 2.75 & 3.14   \\ \hline
    Vocabulary & 2.63 & 2.72   \\ \hline
    Naturalness & 1.88 & 2.15   \\ \hline
    Clarity & 2.63 & 2.86   \\ \hline
    Background Knowledge & 3.63 & 3.43   \\ \hline
    Background Experience & 2.75 & 2.57   \\ \hline
    Agents Behaviors  & 2.63 & 2.72   \\ \hline 
    Conversational Abilities & 3.13 & 3.01   \\ \hline
    Empathic Abilities & 3.13 & 2.86   \\ \hline
  \end{tabular}
\end{center}
\caption{These are the heuristics our system uses to determine what sort of variation to insert.}
\label{table:eval}
\end{table}

We did a two-step evaluation of our system, first on a user to wizard basis, second on a user to agent basis.

\subsection{Wizard of Oz Evaluation}
In order to get an early indication of its real world applicability (e.g. what does an interviewer do and how does empathy play a role), an informal evaluation of the agent begun whilst still under development, using a semi-functional version that was operated via Wizard of Oz set up \cite{hanington2012, dahlback1993}. The interactions were recorded and transcribed for analysis. A second goal of our wizard was to generate a list of generic questions which both elicited good follow-up conversation and were generic enough to apply in different contexts.

\subsection{Agent Evaluation}
Using this refined list of questions we performed a series of interviews with 8 participants (5 male, 3 female). All agent responses were triggered by a human wizard hidden in a separate room with a list of possible responses, manually determining which was best to use. The participants interacted with the wizard for approximately 5 minutes before completing a 16 question post-test questionnaire. The aim of the evaluation survey was to quantitatively assess each user's enjoyment and engagement during the interaction, their perceptions of the naturalness of the interaction, as well as their level of previous knowledge and experience of virtual agents. Users' impressions of the agents likability, depth of vocabulary, quality of non-verbal expression (gestures, facial expressions, timing), clarity of communication, overall conversational abilities as well as her empathic abilities were also assessed. Each metric was defined on the survey and evaluated on a scale of 0-5 ranging from very low to very high, the mean scores $(mean_{1})$ for each metric are included in Table \ref{table:eval}. In our evaluation one of the participants cited their consistently low scores with system latency due to internet problems which continually interrupted the session, we've included the mean scores $(mean_{2})$ without this participant as well.  A qualitative outlet was also provided in order to gather user feedback regarding ideas for the future applications of the agent.

Analysis of participant feedback indicated that participants were generally able to adequately communicate with the agent and answer the questions posed. It was found that the agent evoked a range of responses. The journalist was generally perceived as moderately engaging, with 7 participants reporting the interaction to have been at least moderately to highly enjoyable. However, most participants scored the interaction low on naturalness. Both the verbal and non-verbal communication of the agent received mixed ratings. Only half of the participants found the agent's conversational abilities to be good, whilst the agent's clarity of communication scores particularly varied across participants. The empathy scores reflected that the agent's empathic performance was experienced as being relatively poor. 6 out of 8 participants reported liking virtual agents to some degree, while two were more negative. In terms of suggestions for the potential future applications of the agent, most users suggested that an agent like Alice would be well-suited to the domain of virtual tutoring, or operating as a form of virtual receptionist or social companion, the latter being closest to the planned use of the agent from the ARIA-VALUSPA project's perspective.

\section{Discussion}
The aim of this project was to design and test the efficacy of a virtual agent's ability to elicit information from the user whilst maintaining user interest and emotional engagement. The feedback garnered during the user evaluations showed that the virtual human journalist performed moderately well during the piloting stage,  particularly regarding the duration of sustained conversation held between the agent and user. Most interactions lasted the entirety of the 5 minute allocation without petering out naturally, and were brought to a close by the agent. Therefore our dialogue strategy appears to have been successful in maintaining user engagement.

Scores such as naturalness may have been adversely affected by signaling difficulties which were encountered during testing. The agent did not always articulate words correctly, and some sentences were incoherent, or difficult to hear at times. Additionally some users reported a notable delay in the systems response time, we attribute this to our blackbox TTS/STT (speech-to-text) interface. It should be noted that, as all of the participants were recruited opportunistically during the workshop for the pilot testing stage, all consequently possessed at least some degree of experience and knowledge about virtual agents. While it's evident that a number of shortcomings generally stemmed from technical limitations, the agent's empathic performance was notably poor due to the limited semantic and emotional content of the expressions and gestures available for use. Specifically, user feedback also revealed that the agent's prosody and pronunciation was poor and often contextually-inappropriate. 

Emotional responsiveness is an imperative phenomenon associated with human nature \cite{haslam2006} and the ability to represent functional social bonds \cite{rumble2010}. An agent which attempts to depict a human persona is likely to cause the user to form subconscious expectations as to the agents inherent abilities with respect to empathic understanding. An agent which attempts to depict a human persona is likely to cause the user to form subconscious expectations as to the agents inherent abilities with respect to empathic understanding. In these cases users frequently engage in these over-learned social behaviors, such as politeness and reciprocity, when engaging with traditional computers \cite{Nass2000}, traits which a naive agent cannot properly leverage. Therefore, an agent which puts on this human-like persona but fails to be emotional responsive would immediately be subconsciously recognized as deficient - forming an off-putting environment for the user. Conversely, an agent presented with fewer anthropomorphic cues, subsequently initiating lower user expectations of human-like behavior, could be more easily forgiven for failing to do so. 

Having less anthropomorphic cues would lead to a need for the agent to engage the user in a different way, perhaps by demonstrating other qualities of human nature, such as agency, individuality, cognitive openness and depth of mind \cite{haslam2006}, which may be more convincingly simulated compared to empathy. Whilst an agent depicted as less human could be penalized less for failing to display appropriate emotional cues, such an agent could theoretically be rewarded more for simulating higher cognition, and the ability to engage in deeper levels of abstract thought. Future work may be aimed at testing whether a less humanoid ``philosopher-type" character could be capable of more successfully engaging the user. Such an agent might for instance surprise the user by appearing to be in the midst of an existential crisis, and questioning the meaning of her existence. The agent could be presented to the user in a number forms, varying in terms of politeness in addition to positive and negative personality features. An important point may be for the agent to openly admit to the user to being currently in a learning phase, but also for the agent to demonstrate a genuine eagerness and desire to learn more about what it is to be human.

\section{Future Work}
It was found that Alice failed to engage users emotionally. There are a number of challenges involved in the development of agents that can convincingly simulate empathy and demonstrate emotional responsiveness. Indeed, Ochs, Pelachaud and Sadek \cite{magalie2008} found that users experience agents who respond in an incongruous emotional manner to be more off-putting than agents which do not respond emotionally at all. Although Alice was presented visually to users as being human, it was clear to users that she lacked a deeper understanding of the information presented to her, evident by her lack of emotional responsiveness as a consequence of the limitations of the system.  

The likability scores of the agent reflected that the character Alice was generally liked, but anecdotal feedback from the users who had interacted with her indicated that perhaps her overt politeness may have made her a slightly generic and unmemorable character. Schneiderman \cite{schneiderman2010} argues that an overly friendly and ``humane'' agent should be avoided, as this might make users conclude that the system is more intelligent than it actually is and trust it too much. Particularly experienced users may even experience simulated friendliness as annoying and misleading. Microsoft Bob is an example of such an overtly friendly natural language system that was rejected by the general public \cite{bob}. 

A number of solutions relying on limited anthropomorphic features is already finding its way to the mainstream consumer. Perhaps most notable is the rising tide of intelligent personal assistants, such as Siri. Unlike other voice-controlled services passively awaiting the user command, Siri has been designed to communicate proactively, and even joke with the user, thus strengthening a sense of human personality. However, Siri lacks any form of visual embodied representation, thus limiting a sense of human presence. Similar solutions are currently under development for domestic IoT systems \cite{ivee}. 

A further point of interest would be to examine whether users would be more likely to engage with an embodied virtual agent with more notable or memorable personality features. Future work may be aimed at determining the agent's levels of politeness and displays of overt engagement at which the user retains interest. In other words, it might be useful to determine at what threshold for which politeness and displays of overt engagement does an agent begin to lose user interest. Other future work could examine how users respond to an anti-polite character who is opinionated, argumentative and sometimes rude, and from which point does the user become less amused and engaged and start to find these personality features irritating or disengaging \cite{mckeown2012semaine}. Provisional work was conducted on this in the beginnings of development of a character called Ursula, who was anecdotally found to be more amusing to interact with than Alice.

\section{Conclusion}
Promising first steps were made toward the development of an embodied virtual agent. The agent was capable of utilizing a number of information-eliciting strategies in order to achieve a sustained interaction with a user. We argued that emotional engagement can be improved by enhancing the empathic capabilities of the agent through the integration of automatic emotion recognition and social-signaling software. Additionally, machine learning techniques can automate the process of increasing speech template density. Finally, we hypothesize that users may find a less anthropomorphic agent which exhibits more memorable or unexpected personality features to be more interesting and engaging. 

\section*{Acknowledgment}
The authors would like to thank Professor Michel Valstar and his ARIA-VALUSPA team for granting us access to their development tools. We would also like to extend our thanks to the eNTERFACE workshop organizers for making this project possible.  

\ifCLASSOPTIONcaptionsoff
  \newpage
\fi



%

\bibliography{bib1}
\bibliographystyle{ieeetr}

\end{document}